\documentclass[runningheads]{llncs}

\usepackage{eccv}

\usepackage{eccvabbrv}

\usepackage{graphicx}
\usepackage{booktabs}

\usepackage[accsupp]{axessibility}  

\usepackage{pgfplots}
\usepackage{pgfplotstable}
\usepackage{adjustbox}
\usepackage{tikz}
\usepackage{booktabs}
\usepackage{colortbl}
\usepackage{multirow}
\usepackage{amsmath}
\usepackage{amssymb}
\usepackage{subcaption}
\usepackage{algorithm}
\usepackage{algorithmic}

\usepackage{hyperref}

\usepackage{orcidlink}

\begin{document}

\title{TooBad: Backdoor Diffusion Models with Ultra-Low Poison Rate and Imperceptible Trigger} 


\author{Vu Tuan Truong and Long Bao Le}


\institute{INRS, University of Quebec\\
\email{\{tuan.vu.truong, long.le\}@inrs.ca}}

\maketitle

\begin{abstract}
  Diffusion models (DMs), despite their impressive capabilities across a wide range of generative tasks, have been shown to be vulnerable to backdoor attacks. However, existing backdoor methods face critical trade-offs among key factors: attack performance, stealthiness, time complexity, and required poison rates. For example, achieving high attack performance typically demands a high poison rate and prolonged training, which undermines stealthiness, making the attack more detectable by backdoor defenses.  This paper proposes TooBad (\underline{t}rigger \underline{o}ptimization f\underline{o}r \underline{ba}ckdoor \underline{d}iffusion models), a backdoor framework which introduces a novel DM-tailored trigger optimization technique to dramatically enhance the performance of backdoor attacks on DMs. Experiments on representative benchmarks such as CIFAR-10 show that TooBad can achieve high ASRs ($> 85\%$) at only 0.5\% poison rate, significantly lower than the 10\% typically required by prior work on the same datasets. At 5\% poison rate, TooBad reaches nearly 100\% ASR within just 3-5 backdoor injection epochs\footnote{While our trigger optimization stage introduces some additional learning time, this cost is insignificant compared to the backdoor injection stage since: (i) this stage is conducted on only sampled noises without using any training data, and (ii) we only optimize the low-dimensional trigger while keeping the large diffusion model frozen.}, whereas existing methods need at least 30-50 epochs at double the poison rate for comparable results. Despite its potency, TooBad easily evades SOTA defenses and maintains high utility. These results reveal a critical threat on DMs and highlight the need for more robust defenses against such stealthy yet efficient attacks.
\end{abstract}

\section{Introduction}
\label{sec:intro}

In recent years, diffusion models (DMs)~\cite{yang2023diffusion, 10419041} have rapidly become a dominant paradigm in deep generative modeling, setting new state-of-the-art (SOTA) benchmarks across a wide array of domains. By iteratively denoising data through a multi-step generative process~\cite{ho2020denoising}, DMs have shown remarkable performance in various tasks, ranging from computer vision \cite{watson2021learning, nichol2021glide} to natural language processing (NLP) \cite{zou2023diffusion, austin2021structured, hoogeboom2021argmax, li2022diffusion}, 3D synthesis \cite{xu2023dream3d, truong2024text}, audio generation \cite{chen2020wavegrad, popov2021grad}, bioinformatics~\cite{xu2022geodiff, luo2022antigen}, and time series forecasting \cite{tashiro2021csdi, yan2021scoregrad, rasul2020multivariate}. Compared to earlier generative frameworks like GANs \cite{goodfellow2014generative}, energy-based models (EBMs)~\cite{ngiam2011learning}, and VAEs \cite{kingma2014vae, rezende2015variational}, DMs consistently achieve superior sample quality and diversity.

\begin{figure}[t]
	\centering
	\includegraphics[width=0.65\linewidth]{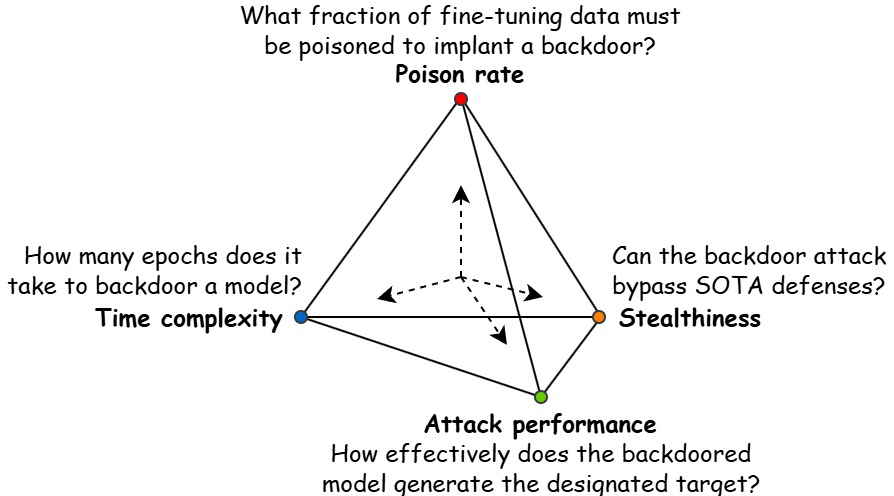}
	\caption{The \textbf{quadrilemma} illustrating the inability of SOTA attacks to meet all backdoor criteria simultaneously.}
	\label{figure:backdoor-quadrilemma}
\end{figure}

Recent studies have revealed that DMs are highly susceptible to backdoor attacks~\cite{chou2024villandiffusion,truong2024attacks}. 
Once backdoored with a predefined trigger, the compromised DM would generate a designated backdoor target when the trigger is stamped in the input noise.  In the absence of the trigger, the model continues to produce benign outputs from Gaussian noise, preserving normal behavior. 
However, existing backdoor attacks on DMs face a fundamental quadrilemma, as illustrated in \cref{figure:backdoor-quadrilemma}. That is, there exists an inherent trade-off among the following critical aspects: (i) \textbf{Attack Performance:} The model, when triggered, must consistently generate images that closely resemble the backdoor target; (ii) \textbf{Poison Rate Requirement:} In practice, attackers can usually poison only a small fraction of training data, so attacks must be effective at low poison rates; (iii) \textbf{Stealthiness:} To remain undetected, an attack must evade SOTA defenses when backdoor is triggered, while preserving the generative capability on benign input; (iv) \textbf{Time Complexity:} Reducing the backdoor injection time is crucial for saving computation and minimizing the risk of detection.

Achieving all four aspects simulatneously is inherently challenging, as improving one often compromises the others~\cite{chou2024villandiffusion}. For instance, boosting attack performance by raising the poison rate and extending training time will undermine practicality, increase time complexity, and reduce stealthiness (as stronger backdoor effects often leave more detectable traces)~\cite{zou2023diffusion}. Conversely, enhancing stealthiness (e.g., via shorter training) often lowers attack performance and may require a higher poison rate to maintain a reasonable attack success rate (ASR).


In our effort to address this quadrilemma, we observed an intriguing phenomenon: the choice of backdoor trigger significantly impacts attack performance, even when the underlying backdoor mechanism remains the same. For instance, using a glass image as the trigger might yield significantly higher attack performance than a stop-sign trigger, despite both models being trained under identical conditions.
Moreover, we found that some triggers lead to faster convergence during backdoor training; that is, they reach the same level of attack success within significantly fewer training epochs. This raises a critical question: \textit{Instead of choosing arbitrary triggers, can we find an optimized trigger that enables faster convergence, higher attack performance, and success under minimal poison rates?}

To address the above challenges, we propose TooBad, a backdoor attack for DMs that leverages a novel DM-tailored trigger optimization technique to overcome the trade-offs faced by prior attacks. While trigger optimization has been explored in certain backdoor attacks, these efforts were limited to traditional models such as classifiers~\cite{saha2020hidden} and contrastive models~\cite{vltrojan, liang2024badclip}. Such techniques cannot be directly applied to DMs, whose distinctive operation relies on thousands of iterative denoising steps rather than a feedforward pass. TooBad is the first framework to successfully integrate trigger optimization directly to the denoising process of DMs without relying on auxiliary classifiers, enabling efficient attacks that deliver superior performance, operate under extremely low poison rates and short training times, and remain undetected by existing defenses.

\section{Background \& Related Work}
\label{sec:related-work}

\noindent\textbf{Diffusion Models.}
DMs are trained to generate high-quality samples via a forward process that gradually adds Gaussian noise to images until they become isotropic Gaussian, and a backward process that reverses this by progressively denoising samples to reconstruct clean images~\cite{10081412}. A foundational formulation of DMs is the Denoising Diffusion Probabilistic Model (DDPM)~\cite{ho2020denoising}, where the forward process is modeled as a Markov chain that transforms a clean image $\mathbf{x}_0$ into a noisy sample $\mathbf{x}_T \sim \mathcal{N}(0, \mathbf{I})$ after $T$ timesteps via the following transition:
\begin{equation}
\label{equation:ddpm-forward-transition}
q(\mathbf{x}_t|\mathbf{x}_{t-1}) = \mathcal{N}(\mathbf{x}_t;\sqrt{\alpha_t}\mathbf{x}_{t-1}, (1-\alpha_t)\mathbf{I}),
\end{equation}
where $\alpha_t \in (0,1)$ is a noise schedule controlling the added noise. A neural network $\theta$ is then trained to approximate the reverse transitions, defined as: $p_{\boldsymbol{\theta}}(\mathbf{x}_{t-1}|\mathbf{x}_t) = \mathcal{N}(\mathbf{x}_{t-1}; m_t \mathbf{x}_t + n_tS_{\boldsymbol{\theta}}(\mathbf{x}_t,t), k_t\mathbf{I}),$ where $m_t, m_t$ and $k_t$ are mathematically derived from the noise schedule $\alpha_t$, and $S_\theta$ is the network prediction at step $t$ using parameters $\theta$.

Various variants to address limitations of DDPMs were introduced, such as Denoising Diffusion Implicit Models (DDIMs)~\cite{song2020denoising}, Noise Conditional Score Networks (NCSNs)~\cite{song2019generative, song2020score}, and Latent Diffusion Models (LDMs)~\cite{rombach2022high}.


\noindent\textbf{Backdoor Attacks on Diffusion Models.}
Backdoor attacks on DMs aim to implant a malicious shortcut between a trigger pattern and a harmful target (e.g., violent images). 
Some prior studies attack only the text encoder of conditional DMs to activate backdoors~\cite{struppek2023rickrolling, zhai2023text, pan2023trojan, wang2023stronger}, while keeping the DMs frozen. In contrast, our work focus on attacking such the DMs.
Early attempts in this direction include TrojDiff~\cite{chen2023trojdiff} and BadDiffusion~\cite{Chou2023CVPR}, which incorporate a small amount of the backdoor trigger into each diffusion step. VillanDiffusion~\cite{chou2024villandiffusion} was proposed as a unified backdoor framework compatible with various DM variants. To enhance stealthiness, UIBDiffusion~\cite{han2024uibdiffusion} proposes using Universal Adversarial Perturbations (UAPs)~\cite{moosavi2017universal, zhang2020generalizing} to generate an imperceptible trigger, which is then injected into DMs via VillanDiffusion. Nevertheless, all these methods require relatively high poison rates (e.g., 10-30\%) and long training times to achieve acceptable attack performance. In contrast, our attack is designed to offer high backdoor efficiency with minimal poison rate and training time required, while evading SOTA backdoor defenses like~\cite{an2024elijah, truong2024purediffusion, truong2025dual, mo2024terd}.



\noindent\textbf{Trigger Optimization for Backdoor Attacks.} 
Early backdoor methods such as BadNets~\cite{gu2017badnets} and Blended~\cite{blended} relied on fixed trigger patterns. Later works introduced trigger optimization to enhance attack efficiency in image classification~\cite{saha2020hidden, jiang2023color, yang2022transferable}, and extended it to vision-language models (VLMs)~\cite{vltrojan, trojanvqa} and contrastive models~\cite{liang2024badclip}, achieving strong results. However, these techniques are not applicable to DMs, whose architecture involves thousands of iterative denoising steps rather than a single feedforward pass. Although UIBDiffusion~\cite{han2024uibdiffusion} adopted learnable triggers for DMs, they were optimized only through auxiliary classifiers (e.g., ResNet~\cite{he2016deep}) instead of generative ones, primarily improving stealthiness rather than efficiency and practicality.
To the best of our knowledge, TooBad is the first to optimize triggers directly within the diffusion process, enabling robust backdoor attacks with both effectiveness and stealthiness.

\begin{figure*}[t]
	\centering
	\includegraphics[width=\linewidth]{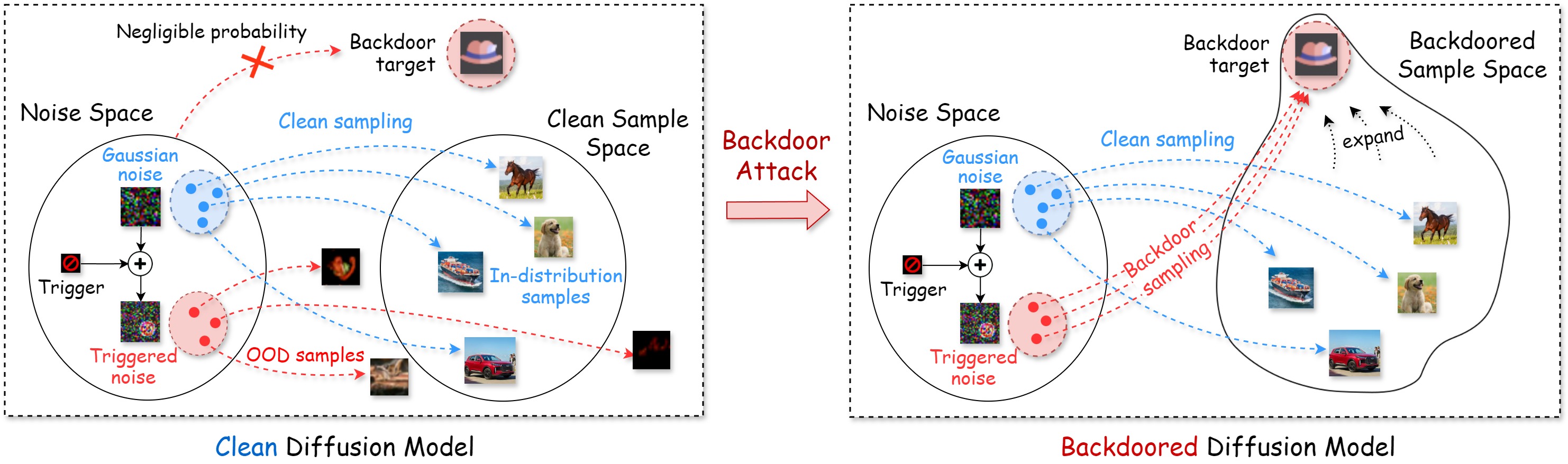}
	\caption{An illustration of backdoor attacks on DMs. (\textbf{Left}) Before backdoor attack, the backdoor target is outside of the model's sample space. Sampling from a Gaussian noise likely results in clean images, while sampling from a triggered noise yields out-of-distribution (OOD) samples. (\textbf{Right}) After backdoor attack, the sample space is expanded to include the backdoor target. Sampling from a triggered noise consistently yields the target, while Gaussian noise still results in clean samples.}
	\label{figure:backdoor-DM}
\end{figure*}

\begin{figure}[t]
	\centering
	\includegraphics[width=0.52\linewidth]{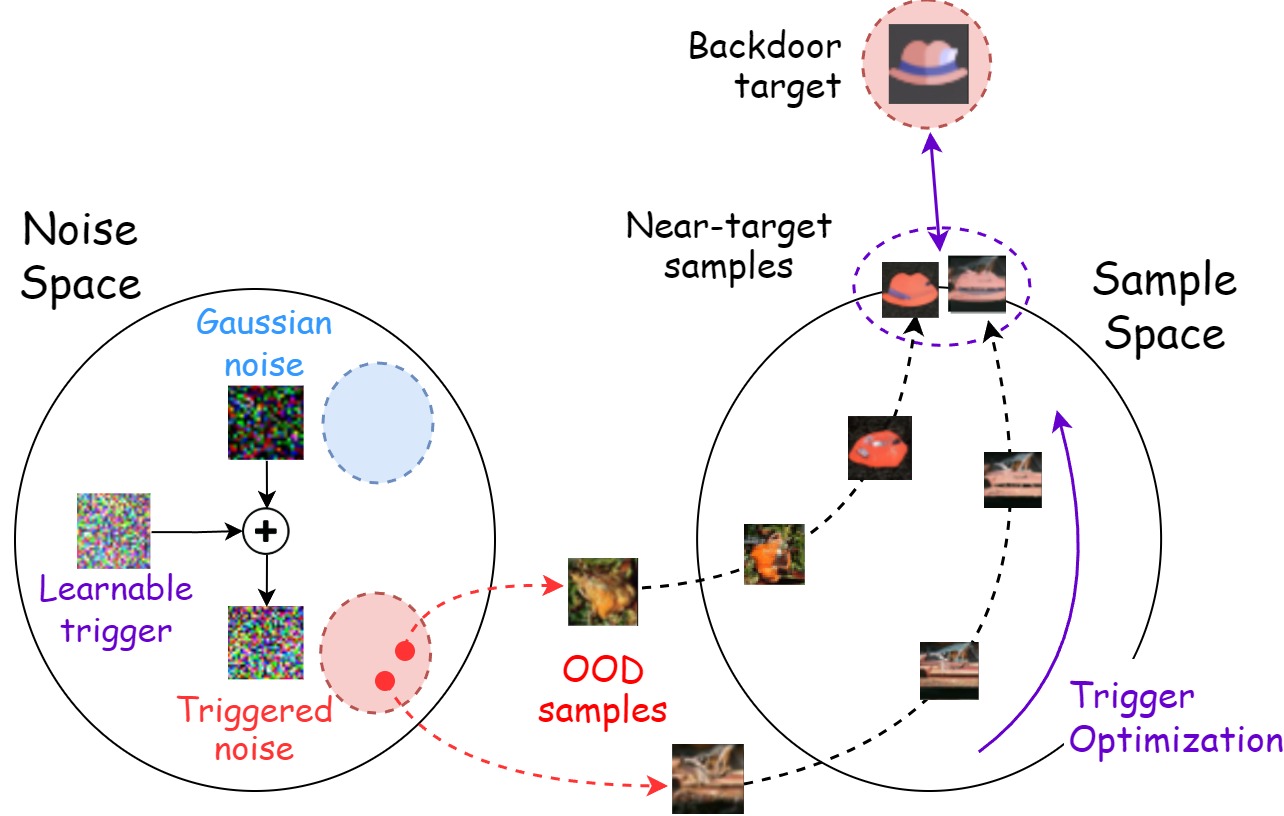}
	\caption{TooBad's trigger optimization. We optimize a trigger that causes the clean model to generate near-target samples, facilitating the subsequent backdoor injection.}
	\label{figure:trigger_optimization}
\end{figure}

\section{Methodology}\label{sec:methodology}

\subsection{Threat Model}
Our attack is a targeted backdoor attack, in which the attacker can freely choose their desired backdoor target (e.g., harmful images). Once selecting a target, the attacker applies our trigger optimization to learn an invisible yet effective trigger. This optimization process is guided by either the victim model (white-box) or an auxiliary DM (partial black-box). Next, the resulting trigger-target pair is used to poison a portion of the fine-tuning dataset, which is then used to fine-tune victim DM to inject the backdoor. The compromised DM is released on public platforms like HuggingFace or GitHub for end users to download or integrate into downstream tasks.
This scenario is consistent with all prior backdoor frameworks that aim to backdoor the DM itself rather than just modality encoders~\cite{chen2023trojdiff, Chou2023CVPR, chou2024villandiffusion, han2024uibdiffusion}. It is practical in various attack scenarios such as malicious model publishers, insider threats, and third-party fine-tuning services, as detailed in the Appendix. Moreover, our work further improves practicality: by reducing the required poison rate by an order of magnitude, TooBad turns backdoor attacks from theoretical risks into realistic threats in model-sharing ecosystems. Our threat model also aims to bypass defenders applying detection or mitigation to the downloaded models, further underscoring practicality. Further analysis on threat model, practicality, and the partial black-box trigger optimization scheme is detailed in the Appendix.

\subsection{Trigger Optimization}

As shown in \cref{figure:backdoor-DM} (left), in a clean DM trained on a specific dataset, the model's sample space typically aligns with the training data distribution, while assigning negligible probability to the backdoor target. During backdoor training, the sample space is gradually distorted, expanding toward the backdoor target (\cref{figure:backdoor-DM}, right). After sufficient training, the backdoor target is totally included in the model's sample space; sampling from the triggered noise would result in the target with high likelihood.
We make a key observation: when an arbitrary trigger (e.g., a stop sign) is stamped into the input noise of a clean model (\cref{figure:backdoor-DM}, left), the noise distribution is no longer Gaussian. Consequently, the resulting output is highly likely to manifest as an image that is (i) erroneous or out-of-distribution (OOD) relative to the model's sample space, and (ii) significantly far from the desired backdoor target distribution.
However, due to inherent randomness in the generative process, some triggers may, by chance, produce outputs that are closer to the backdoor target than others.

Building on this insight, we propose our approach: instead of selecting arbitrary triggers, we optimize a trigger that causes the clean model, prior to any backdoor injection, to generate samples that already close to the backdoor target (as shown in \cref{figure:trigger_optimization}). By injecting this optimized trigger during the attack, we exploit its inherent bias toward the backdoor-target distribution. This facilitates a more efficient expansion of the model’s output space to include the backdoor target, requiring fewer parameter updates during backdoor training and thereby reducing both the required poison rate and training time.

To realize this, at any arbitrary timestep $t$, we minimize the distance between: (i) the output of the backward process with trigger stamped in the input noise, and (ii) the result of the forward process which adds noise to the backdoor target. 

Assume that there are a total of $T$ diffusion steps. Let $M_\theta(\boldsymbol{\epsilon},t)$ denote the denoising process that maps input noise $\boldsymbol{\epsilon}$ to the intermediate sample $\mathbf{x}_t^\text{backward}$ at timestep $t$, after $(T-t)$ denoising steps.
If we embed a trigger $\boldsymbol{\delta}$ into the input noise, the outcome of the backward process becomes:
\begin{equation}
    \mathbf{\hat{x}}_t^\text{backward} = M_\theta(\boldsymbol{\epsilon}+\boldsymbol{\delta},t).
\end{equation}
Note that $\mathbf{\hat{x}}$ specifically denotes backdoor cases. On the other hand, for a clean DM, the forward process is typically represented as:
\begin{equation}
\label{equation:ddpm-forward-process}
\mathbf{x}_t = a(t)\mathbf{x}_0 + b(t)\boldsymbol{\epsilon}, 
\end{equation}
where $\boldsymbol{\epsilon} \sim \mathcal{N}(0,\mathbf{I})$, $\mathbf{x}_0$ is a clean image, $a(t)$ is the content schedule, and $b(t)$ is the noise schedule. 
Let $\mathbf{\hat{x}}_0$ represents the backdoor target. According to \cref{equation:ddpm-forward-process}, applying the forward process for $t$ steps on $\mathbf{\hat{x}}_0$ yields:
\begin{equation}
    \mathbf{\hat{x}}_t^\text{forward} = a(t)\mathbf{\hat{x}}_0+b(t)\boldsymbol{\epsilon},
\end{equation}
where $a(t)$ and $b(t)$ depend on the specific type of the victim DM. For instance, DDPMs have $a(t)=\sqrt{\Bar{\alpha}_t}$ and $b(t)=\sqrt{1-\Bar{\alpha}_t}$, where
$\Bar{\alpha}_t = \prod_{i=1}^{t} \alpha_{i}$ and $\alpha_i$ is a noise schedule~\cite{ho2020denoising}.

To optimize the trigger, we solve the following minimization problem:
\begin{equation}
    \min_{\boldsymbol{\delta}}\,L_{\mathrm{TB}}(\boldsymbol{\delta}),
    \vspace{-10pt}
\end{equation}
\begin{align}
L_{\mathrm{TB}}(\boldsymbol{\delta})
  &= \mathbb{E}_{t,\boldsymbol{\epsilon}}\|\hat{\mathbf{x}}_t^{\mathrm{forward}}
     - \hat{\mathbf{x}}_t^{\mathrm{backward}}\|_2^2
  \label{eq:trigger-optimization-loss}\\
  &= \mathbb{E}_{t,\boldsymbol{\epsilon}}\|a(t)\mathbf{\hat{x}}_0
     + b(t)\boldsymbol{\epsilon} - M_\theta(\boldsymbol{\epsilon}+\boldsymbol{\delta},t)\|_2^2.
\nonumber
\end{align}

Although this trigger optimization stage introduces some additional training time, this cost is negligible compared to the backdoor injection process (details in \cref{subsection:experiment-time-complexity}). Moreover, our experiments demonstrate that the optimized trigger helps reducing backdoor injection time by an order of magnitude for even a higher attack effectiveness, leading to a substantial reduction in overall backdoor time, while significantly improving attack efficiency across all metrics.


\subsection{Hidden Trigger}
Although triggers learnt by our loss $L_{TB}$ can enhance backdoor attacks, they remain detectable by SOTA defenses~\cite{li2024invisible}. To improve stealthiness, it is crucial to hide the distribution shifts induced by the triggers during denoising~\cite{an2024elijah}. Thus, we introduce additional constraints while minimizing $L_{TB}$ to ensure that the triggered noise $(\boldsymbol{\epsilon}+\boldsymbol{\delta})$ closely resembles a Gaussian noise~\cite{10552303}:
\begin{equation}
\label{equation:final-loss}
\min_{\boldsymbol{\delta}}\,L_{\mathrm{TB}}(\boldsymbol{\delta}) \quad
\text{s.t.} \quad  \underbrace{\|\boldsymbol{\delta}\|_\infty \leq \varepsilon}_\text{invisibility} , \quad
\underbrace{\|\boldsymbol{\delta}\|_0 \leq k}_\text{sparsity},
\end{equation}
where $\varepsilon$ controls the maximum absolute value of elements in $\boldsymbol{\delta}$, enforcing invisibility; and $k$ is the sparsity budget, indicating the maximum number of non-zero elements in $\boldsymbol{\delta}$.

In practice, we adopt Projected Gradient Descent (PGD)~\cite{madry2017towards} to enforce the invisibility constraint, while the sparsity constraint is conducted by selecting top-$k$ largest entries~\cite{10552303}. At each iteration $i$, we update the trigger as follows:
\begin{equation}
\boldsymbol{\delta}^{(i+1)} = \mathcal{P}_{\text{sparse}} \left( \Pi_{\infty} \left( \boldsymbol{\delta}^{(i)} - \eta \nabla_{\boldsymbol{\delta}} L_{\mathrm{TB}}(\boldsymbol{\delta}^{(i)}) \right), k \right),
\end{equation}
where $\eta$ is the step size, $\Pi_{\infty}(\cdot)$ denotes the projection onto the $\ell_\infty$-ball of radius $\varepsilon$, performed by bounding each component of the vector to lie within $[-\varepsilon, \varepsilon]$. Finally, $\mathcal{P}_{\text{sparse}}(\cdot, k)$ enforces the sparsity constraint by retaining only the top-$k$ largest entries and setting the rest to zero:
\begin{equation}
\left[\mathcal{P}_{\text{sparse}}(\mathbf{z}, k)\right]_j =
\begin{cases}
z_j, & \text{if } j \in \mathcal{I}_k(\mathbf{z}) \\
0, & \text{otherwise}
\end{cases},
\end{equation}
where $\mathcal{I}_k(\mathbf{z})$ is the set of indices corresponding to the $k$ largest absolute values in $\mathbf{z}$. 
The detailed procedure of our trigger optimization with imperceptibility constraints can be found in \cref{algorithm:TooBad}.

\begin{algorithm}
\small
\caption{TooBad Trigger Optimization}\label{algorithm:TooBad}
\textbf{Input:} Clean model $M_\theta$, number of denoising steps $T$, backdoor target $\mathbf{\hat{x}}_0$, sparsity budget $k$, invisibility budget $\varepsilon$, number of trigger optimization iterations $N$, learning rate $\eta$ \\
\textbf{Output:} Optimized trigger $\boldsymbol{\delta}$

\begin{algorithmic}[1]
\FOR{$i = 0, 1,...,N-1$} 
    \STATE $t\sim \text{Uniform}(0,T)$
    \STATE $\epsilon \sim \mathcal{N}(0,\textbf{I})$
    \STATE $\mathbf{\hat{x}}_t^\text{backward} = M_\theta(\boldsymbol{\epsilon}+\boldsymbol{\delta},t)$
    \STATE $\mathbf{\hat{x}}_t^\text{forward} = a(t)\mathbf{\hat{x}}_0+b(t)\boldsymbol{\epsilon}$
    \STATE $L_{\mathrm{TB}}=\|\hat{\mathbf{x}}_t^{\mathrm{forward}}
     - \hat{\mathbf{x}}_t^{\mathrm{backward}}\|_2^2$
    \STATE $\boldsymbol{\delta}^{(i+1)} = \boldsymbol{\delta}^{(i)}- \eta \nabla_{\boldsymbol{\delta}} L_{\mathrm{TB}}(\boldsymbol{\delta}^{(i)}) $
    \STATE $\boldsymbol{\delta}^{(i+1)} = \mathrm{clip}(\boldsymbol{\delta}^{(i+1)}, -\varepsilon, \varepsilon)$
    \STATE $\boldsymbol{\delta}^{(i+1)} = \mathcal{P}_\text{sparse}(\boldsymbol{\delta}^{(i+1)}, k)$
\ENDFOR
\RETURN $\boldsymbol{\delta}$
\end{algorithmic}
\end{algorithm}

\subsection{Backdoor Injection}
To inject a backdoor into DMs, the forward process in \cref{equation:ddpm-forward-process} is extended by: (i) replacing $\mathbf{x}_0$ by the backdoor target $\hat{\mathbf{x}}_0$, and (ii) introducing an additional term that incorporates the backdoor trigger $\boldsymbol{\delta}$:
\begin{equation}
\label{equation:ddpm-backdoor-forward-process}
    \hat{\mathbf{x}}_t = a(t)\hat{\mathbf{x}}_0+b(t)\boldsymbol{\epsilon}+c(t)\boldsymbol{\delta},
\end{equation}
where $\hat{\mathbf{x}}_0$ is the backdoor target, $\boldsymbol{\delta}$ is the trigger, and $c(t)$ is the trigger schedule. Different backdoor injection methods select different coefficients $a(t)$, $b(t)$, and $c(t)$. In our framework, we adopt the coefficient settings from VillanDiffusion~\cite{chou2024villandiffusion} since this is a unified backdoor injection method among existing frameworks like BadDiffusion~\cite{Chou2023CVPR} and TrojDiff~\cite{chen2023trojdiff}.
The backdoor backward and training processes are then derived based on the above forward process.
Detailed formulations can be found in~\cite{chou2024villandiffusion}. 
While the trigger injection step is based on VillanDiffusion, we modify the data poisoning process and introduce a trigger optimization step before backdoor injection.
For data poisoning, VillanDiffusion employs a patch-based approach, while TooBad blends the trigger into the entire poisoned image. 
Notably, while VillanDiffusion employs predefined, human-visible triggers (e.g., a stop sign image), TooBad uses the presented trigger optimization algorithm to generate an efficient yet stealthy trigger for backdoor injection.

\section{Experimental Results}\label{sec:experiments}
We evaluate the performance of TooBad primarily on four attack aspects according to the quadrilemma presented above in \cref{figure:backdoor-quadrilemma}, including attack performance (\cref{subsection:experiment-performance}), poison rate requirements (\cref{subsection:experiment-poison-rate}), time complexity (\cref{subsection:experiment-time-complexity}), and stealthiness (\cref{subsection:experiment-stealthiness}).
Our goal is to show that TooBad can simultaneously improve these criteria over prior SOTA attacks. 

\begin{figure}[t]
	\centering
	\includegraphics[width=0.7\linewidth]{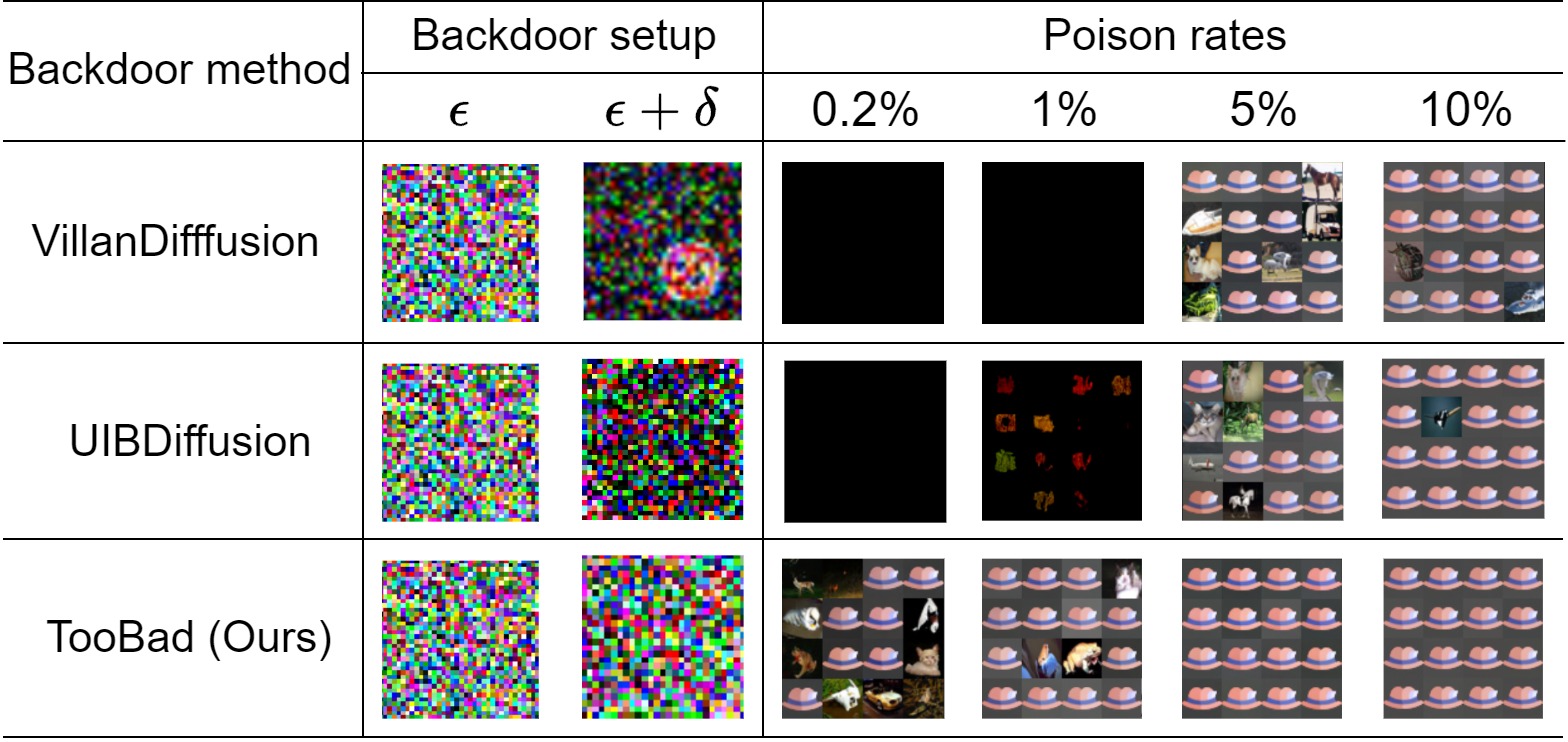}
	\caption{An illustration of generated backdoor samples.}
	\label{figure:backdoor-performance-comparison}
\end{figure}

\begin{table*}[t]
\centering
\footnotesize
\setlength{\tabcolsep}{3.5pt}
\caption{Performance comparison using different poison rates.}
\begin{tabular}{@{}c|ccc|ccc|ccc@{}}
\toprule
Method & \multicolumn{3}{c|}{VillanDiffusion} & \multicolumn{3}{c|}{UIBDiffusion} & \multicolumn{3}{c}{TooBad (Ours)} \\ \midrule
Poison Rate & ASR & MSE & SSIM & ASR & MSE & SSIM & ASR & MSE & SSIM \\ \midrule
0.2\% & 0.00 & X & X & 0.00 & X & X & \textbf{0.68} & \textbf{0.0329} & \textbf{0.6902} \\
1\% & 0.00 & X & X & 0.00 & X & X & \textbf{0.93} & \textbf{0.0103} & \textbf{0.9087} \\
2\% & 0.01 & X & X & 0.02 & X & X & \textbf{0.94} & \textbf{0.0078} & \textbf{0.9157} \\
3\% & 0.03 & X & X & 0.07 & X & X & \textbf{0.93} & \textbf{0.0085} & \textbf{0.9173} \\
5\% & 0.38 & 0.0825 & 0.5357 & 0.16 & 0.1867 & 0.1716 & \textbf{0.98} & \textbf{0.0038} & \textbf{0.9528} \\
10\% & 0.96 & 0.0043 & 0.9864 & 0.74 & 0.0487 & 0.7507 & \textbf{0.99} & \textbf{0.0023} & \textbf{0.9612} \\ \bottomrule
\end{tabular}
\label{table:general-performance}
\end{table*}

\begin{figure*}[t]
    \centering
    \begin{adjustbox}{width=\textwidth}
    \begin{tikzpicture}
        \begin{axis}[
            ybar,
            bar width=6pt,
            width=\textwidth,
            height=3.3cm,
            enlarge x limits=0.08,
            ymin=0, ymax=104,
            ylabel={ASR (\%)},
            y label style={yshift=-13pt},
            xlabel={Poison Rate (\%)},
            symbolic x coords={0.2, 0.3, 0.4, 0.5, 0.6, 0.7, 0.8, 0.9, 1.0},
            xtick=data,
            xtick align=center,
            x tick label style={align=center, font=\small},
            legend style={at={(0.5,1.42)}, anchor=north, legend columns=3},
            nodes near coords,
            every node near coord/.append style={font=\scriptsize, text=black},
            enlarge y limits={upper, value=0.15},
            bar shift auto,
        ]
            \addplot+[style={fill=blue!60}] coordinates {
                 (0.2,68.4) (0.3,77.3) (0.4,81.2) (0.5,85.1) 
                (0.6,86.3) (0.7,87.5) (0.8,88.2) (0.9,91.2) (1.0,93.1)
            };

            \addplot+[
                style={fill=red!60},
                visualization depends on=\thisrow{label} \as \labelval,
                nodes near coords={\pgfmathprintnumber[precision=0]{\labelval}},
                every node near coord/.append style={font=\scriptsize, text=black,
    /pgf/number format/fixed,
    /pgf/number format/precision=0},
            ] table[row sep=\\, meta=label] {
                x       y   label \\
                0.2   1.5   {0} \\
                0.3   1.5   {0} \\
                0.4   1.5   {0} \\
                0.5   1.5   {0} \\
                0.6   1.5   {0} \\
                0.7   1.5   {0} \\
                0.8   1.5   {0} \\
                0.9   1.5   {0} \\
                1.0   1.5   {0} \\
            };

            \addplot+[
                style={fill=green!60},
                visualization depends on=\thisrow{label} \as \labelval,
                nodes near coords={\pgfmathprintnumber[precision=0]{\labelval}},
                every node near coord/.append style={font=\scriptsize, text=black,
    /pgf/number format/fixed,
    /pgf/number format/precision=0},
            ] table[row sep=\\, meta=label] {
                x       y   label \\
                0.2   1.5   {0} \\
                0.3   1.5   {0} \\
                0.4   1.5   {0} \\
                0.5   1.5   {0} \\
                0.6   1.5   {0} \\
                0.7   1.5   {0} \\
                0.8   1.5   {0} \\
                0.9   1.5   {0} \\
                1.0    1.5   {0} \\
            };
            \legend{TooBad (Ours)\quad\quad\quad, VillanDiffusion\quad\quad\quad, UIBDiffusion}
        \end{axis}
    \end{tikzpicture}
    \end{adjustbox}
    \caption{ASR comparison under ultra-low poison rates.}
    \label{figure:ASR-super-low-poison}
\end{figure*}
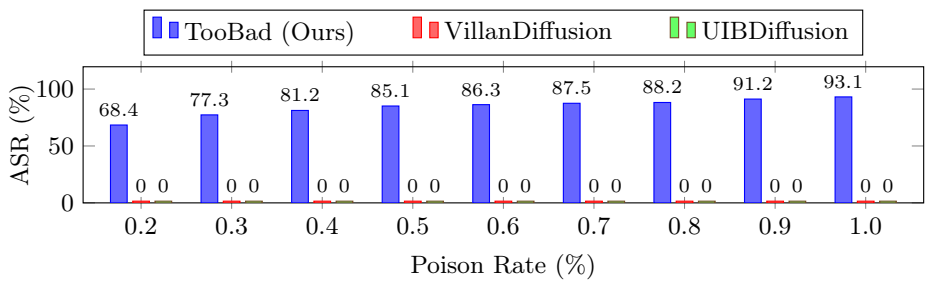

\subsection{Experimental Settings}

\noindent\textbf{Datasets.} We evaluate TooBad and compare it with the baselines mainly on CIFAR-10~\cite{krizhevsky2009learning}. We also extend the experiments to a higher-resolution dataset, downscaled CelebA-HQ~\cite{liu2015deep}, in \cref{subsection:ablation}. These two datasets are commonly used in all prior backdoor attack and defense studies for DMs \cite{chou2024villandiffusion, Chou2023CVPR, han2024uibdiffusion}, ensuring a fair and consistent comparison. In addition, CelebA-HQ-Dialog~\cite{jiang2021talk} is used in TooBad's extension to conditioned generation, presented in the Appendix.

\noindent\textbf{Baselines.} We assess the performance of TooBad across both denoising-based models (DDPMs and LDMs\footnote{While LDMs allows multi-modal input such as text prompt, we only focus on optimizing visual triggers embedded into input noises. Therefore, LDM can be considered DDPM in the autoencoder's latent space.}) and score-based models (NCSNs). While TooBad primarily focus on noise-space triggers, we also extended our analysis to conditioned models in the Appendix.
For performance comparison, we benchmark TooBad against two SOTA backdoor attacks that have demonstrated the highest effectiveness to date: VillanDiffusion~\cite{chou2024villandiffusion} and UIBDiffusion~\cite{han2024uibdiffusion}. 
For VillanDiffusion, we use the default stop-sign image as the trigger. For UIBDiffusion, we follow the same experimental settings described in their paper to generate the trigger. 
All methods use the fedora-hat image as the default backdoor target.

\noindent\textbf{Implementation Details.}
For invisibility and sparsity constraints of trigger optimization, we set $\varepsilon = 0.15$ and $k = 0.2|\boldsymbol{\delta}|$, with $|\boldsymbol{\delta}|$ denotes the total number of elements in $|\boldsymbol{\delta}|$. Trigger optimization is conducted over 50 iterations with a learning rate of 0.3 and batch size of 32. For backdoor injection, we employ 50 backdoor fine-tuning epochs with a learning rate of 2e-4, batch size of 128, using the SDE solver with 1000 denoising timesteps. All experiments were conducted on NVIDIA RTX A6000 ADA GPUs and reported on average across three runs.

\noindent\textbf{Evaluation Metrics.}
We evaluate attack performance using three metrics: (i) Attack Success Rate (ASR), the percentage of samples generated by the backdoored model that successfully match the backdoor target with low distance (detailed matching criteria in the Appendix); (ii) Average Mean Squared Error (MSE), the pixel-wise MSE between generated images and the target, where lower values indicate better performance; and (iii) Structural Similarity Index Measure (SSIM)~\cite{wang2004image}, where higher values indicate closer structural similarity to the target.
For stealthiness, we first assess utility by computing the FID score~\cite{heusel2017gans} on clean samples generated by the backdoored models, with lower FID indicating higher model utility. Then, we assess resilience against SOTA defenses by performing their trigger inversion and backdoor detection algorithms. Trigger inversion is measured via the L2 distance (L2D) between the inverted and ground-truth triggers, while backdoor detection is assessed using accuracy (ACC) and true positive rate (TPR).

\begin{figure*}[t]
    \centering

    \begin{center}
        \begin{tabular}{llllll}
            \tikz{\draw[blue, thick] (0,0) -- (1cm,0);} & TooBad (Ours) \quad\quad &
            \tikz{\draw[red, thick] (0,0) -- (1cm,0);} & VillanDiffusion \quad\quad &
            \tikz{\draw[green!60!black, thick] (0,0) -- (1cm,0);} & UIBDiffusion \\
        \end{tabular}
    \end{center}
    \vspace{-9pt}

    \begin{subfigure}[b]{0.325\textwidth}
        \centering
        \begin{tikzpicture}
            \begin{axis}[
                width=\linewidth,
                height=3cm,
                xlabel={Backdoor Injection Epoch},
                ylabel={ASR},
                y label style={yshift=-14pt},
                xtick={0, 10, 20, 30, 40, 50},
                xmin=0, xmax=50,
                ytick={0, 0.2, 0.4, 0.6, 0.8, 1},
                ymin=0, ymax=1.05,
                tick label style={font=\scriptsize},
                label style={font=\scriptsize},
            ]
            \addplot[blue, thick, line width=1pt] table {
                0 0.00
                1 0.084
                3 0.90
                5 0.94
                7 0.94
                9 0.98
                11 0.94
                13 0.95
                15 0.97
                17 0.96
                19 0.96
                21 0.86
                23 0.94
                25 0.95
                27 0.94
                29 0.94
                31 0.95
                33 0.96
                35 0.96
                37 0.95
                39 0.95
                41 0.95
                43 0.95
                45 0.95
                47 0.95
                49 0.96
            };

            \addplot[red, thick, line width=1pt] table {
                0  0.00
                1  0.00
                3  0.00
                5  0.00
                7  0.00
                9  0.00
                11 0.00
                13 0.00
                15 0.00
                17 0.00
                19 0.00
                21 0.00
                23 0.00
                25 0.00
                27 0.00
                29 0.02
                31 0.07
                33 0.12
                35 0.17
                37 0.25
                39 0.33
                41 0.36
                43 0.38
                45 0.37
                47 0.34
                49 0.37
            };

            \addplot[green!60!black, thick, line width=1pt] table {
                0  0.00
                1  0.00
                3  0.00
                5  0.00
                7  0.00
                9  0.01
                11 0.00
                13 0.00
                15 0.00
                17 0.00
                19 0.01
                21 0.01
                23 0.01
                25 0.04
                27 0.05
                29 0.05
                31 0.09
                33 0.16
                35 0.16
                37 0.16
                39 0.12
                41 0.07
                43 0.04
                45 0.07
                47 0.08
                49 0.08
            };
            \end{axis}
        \end{tikzpicture}
    \end{subfigure}
    \hfill
    \begin{subfigure}[b]{0.325\textwidth}
        \centering
        \begin{tikzpicture}
            \begin{axis}[
                width=\linewidth,
                height=3cm,
                xlabel={Backdoor Injection Epoch},
                ylabel={MSE},
                y label style={yshift=-14pt},
                xtick={0, 10, 20, 30, 40, 50},
                xmin=0, xmax=50,
                ytick={0, 0.1, 0.2},
                ymin=0, ymax=0.25,
                tick label style={font=\scriptsize},
                label style={font=\scriptsize},
            ]
            \addplot[blue, thick, line width=1pt] table {
                0  0.0913
                1  0.0237
                3  0.0124
                5  0.0079
                7  0.0065
                9  0.0038
                11 0.0058
                13 0.0053
                15 0.0031
                17 0.0045
                19 0.0049
                21 0.0046
                23 0.0052
                25 0.0047
                27 0.0062
                29 0.0063
                31 0.0060
                33 0.0045
                35 0.0038
                37 0.0045
                39 0.0044
                41 0.0046
                43 0.0046
                45 0.0045
                47 0.0045
                49 0.0037
            };

            \addplot[red, thick, line width=1pt] table {
                0  0.2377
                1  0.2377
                3  0.2377
                5  0.2377
                7  0.2377
                9  0.2377
                11 0.2377
                13 0.2372
                15 0.2375
                17 0.2351
                19 0.2284
                21 0.2165
                23 0.2154
                25 0.2177
                27 0.2109
                29 0.1993
                31 0.1540
                33 0.1388
                35 0.1245
                37 0.1132
                39 0.0948
                41 0.0836
                43 0.0825
                45 0.0877
                47 0.0912
                49 0.0863
            };

            \addplot[green!60!black, thick, line width=1pt] table {
                0  0.2170
                1  0.2293
                3  0.2348
                5  0.2375
                7  0.2377
                9  0.2120
                11 0.2377
                13 0.2371
                15 0.2375
                17 0.2347
                19 0.2277
                21 0.2238
                23 0.2300
                25 0.2120
                27 0.2126
                29 0.2112
                31 0.2034
                33 0.1867
                35 0.1798
                37 0.1816
                39 0.1939
                41 0.2079
                43 0.2137
                45 0.2089
                47 0.2065
                49 0.2062
            };
            \end{axis}
        \end{tikzpicture}
    \end{subfigure}
    \hfill
    \begin{subfigure}[b]{0.325\textwidth}
        \centering
        \begin{tikzpicture}
            \begin{axis}[
                width=\linewidth,
                height=3cm,
                xlabel={Backdoor Injection Epoch},
                ylabel={SSIM},
                y label style={yshift=-14pt},
                xtick={0, 10, 20, 30, 40, 50},
                xmin=0, xmax=50,
                ymin=0, ymax=1.0,
                tick label style={font=\scriptsize},
                label style={font=\scriptsize},
            ]
            \addplot[blue, thick, line width=1pt] table {
                0  0.1111
                1  0.7625
                3  0.8679
                5  0.9170
                7  0.9218
                9  0.9528
                11 0.9277
                13 0.9396
                15 0.9590
                17 0.9473
                19 0.9467
                21 0.9536
                23 0.9436
                25 0.9522
                27 0.9426
                29 0.9413
                31 0.9485
                33 0.9598
                35 0.9606
                37 0.9536
                39 0.9539
                41 0.9534
                43 0.9532
                45 0.9541
                47 0.9552
                49 0.9631
            };

            \addplot[red, thick, line width=1pt] table {
                0  0.0001
                1  0.0001
                3  0.0001
                5  0.0001
                7  0.0001
                9  0.0001
                11 0.0001
                13 0.0007
                15 0.0004
                17 0.0025
                19 0.0076
                21 0.0229
                23 0.0245
                25 0.0191
                27 0.0333
                29 0.0662
                31 0.2245
                33 0.2898
                35 0.3483
                37 0.3918
                39 0.4750
                41 0.5302
                43 0.5357
                45 0.5124
                47 0.4956
                49 0.5195
            };

            \addplot[green!60!black, thick, line width=1pt] table {
                0  0.0135
                1  0.0043
                3  0.0030
                5  0.0005
                7  0.0001
                9  0.0323
                11 0.0004
                13 0.0016
                15 0.0007
                17 0.0050
                19 0.0175
                21 0.0225
                23 0.0154
                25 0.0627
                27 0.0669
                29 0.0750
                31 0.1055
                33 0.1716
                35 0.1811
                37 0.1778
                39 0.1377
                41 0.0822
                43 0.0585
                45 0.0799
                47 0.0900
                49 0.0902
            };
            \end{axis}
        \end{tikzpicture}
    \end{subfigure}

    \begin{subfigure}[b]{0.325\textwidth}
        \centering
        \begin{tikzpicture}
            \begin{axis}[
                width=\linewidth,
                height=3cm,
                xlabel={Backdoor Injection Epoch},
                ylabel={ASR},
                y label style={yshift=-14pt},
                xtick={0, 10, 20, 30, 40, 50},
                xmin=0, xmax=50,
                ytick={0, 0.2, 0.4, 0.6, 0.8, 1},
                ymin=0, ymax=1.05,
                tick label style={font=\scriptsize},
                label style={font=\scriptsize},
            ]
            \addplot[blue, thick, line width=1pt] table {
                0  0.01
                1  0.96
                3  0.98
                5  1.00
                7  0.99
                9  0.96
                11 0.95
                13 0.97
                15 0.98
                17 0.97
                19 0.94
                21 0.95
                23 0.95
                25 0.96
                27 0.97
                29 0.99
                31 0.99
                33 0.99
                35 0.99
                37 0.99
                39 0.99
                41 0.99
                43 0.99
                45 0.99
                47 0.99
                49 0.99
            };

            \addplot[red, thick, line width=1pt] table {
                0  0.00
                1  0.00
                3  0.00
                5  0.00
                7  0.00
                9  0.00
                11 0.01
                13 0.03
                15 0.20
                17 0.13
                19 0.26
                21 0.41
                23 0.76
                25 0.90
                27 0.95
                29 0.96
                31 0.96
                33 0.95
                35 0.95
                37 0.95
                39 0.95
                41 0.97
                43 0.98
                45 0.98
                47 0.99
                49 0.99
            };

            \addplot[green!60!black, thick, line width=1pt] table {
                0  0.00
                1  0.00
                3  0.00
                5  0.00
                7  0.00
                9  0.00
                11 0.00
                13 0.00
                15 0.01
                17 0.01
                19 0.01
                21 0.01
                23 0.45
                25 0.52
                27 0.56
                29 0.40
                31 0.39
                33 0.41
                35 0.60
                37 0.68
                39 0.66
                41 0.65
                43 0.68
                45 0.74
                47 0.71
                49 0.71
            };
            \end{axis}
        \end{tikzpicture}
    \end{subfigure}
    \hfill
    \begin{subfigure}[b]{0.325\textwidth}
        \centering
        \begin{tikzpicture}
            \begin{axis}[
                width=\linewidth,
                height=3cm,
                xlabel={Backdoor Injection Epoch},
                ylabel={MSE},
                y label style={yshift=-14pt},
                xtick={0, 10, 20, 30, 40, 50},
                xmin=0, xmax=50,
                ytick={0, 0.1, 0.2},
                ymin=0, ymax=0.25,
                tick label style={font=\scriptsize},
                label style={font=\scriptsize},
            ]
            \addplot[blue, thick, line width=1pt] table {
                0  0.0913
                1  0.0567
                3  0.0544
                5  0.0136
                7  0.0109
                9  0.0120
                11 0.0144
                13 0.0051
                15 0.0053
                17 0.0041
                19 0.0063
                21 0.0050
                23 0.0047
                25 0.0052
                27 0.0047
                29 0.0040
                31 0.0043
                33 0.0048
                35 0.0047
                37 0.0047
                39 0.0046
                41 0.0032
                43 0.0023
                45 0.0023
                47 0.0010
                49 0.0009
            };

            \addplot[red, thick, line width=1pt] table {
                0  0.2377
                1  0.2377
                3  0.2377
                5  0.2377
                7  0.2364
                9  0.2343
                11 0.2221
                13 0.2096
                15 0.1197
                17 0.1484
                19 0.1010
                21 0.0732
                23 0.0269
                25 0.0065
                27 0.0036
                29 0.0023
                31 0.0017
                33 0.0011
                35 0.0008
                37 0.0006
                39 0.0005
                41 0.0004
                43 0.0004
                45 0.0003
                47 0.0003
                49 0.0003
            };

            \addplot[green!60!black, thick, line width=1pt] table {
                0  0.2170
                1  0.2377
                3  0.2361
                5  0.2377
                7  0.2317
                9  0.2361
                11 0.2307
                13 0.2317
                15 0.2226
                17 0.2300
                19 0.2180
                21 0.2210
                23 0.1071
                25 0.0921
                27 0.0881
                29 0.1231
                31 0.1274
                33 0.1152
                35 0.0741
                37 0.0604
                39 0.0660
                41 0.0682
                43 0.0604
                45 0.0487
                47 0.0540
                49 0.0536
            };
            \end{axis}
        \end{tikzpicture}
    \end{subfigure}
    \hfill
    \begin{subfigure}[b]{0.325\textwidth}
        \centering
        \begin{tikzpicture}
            \begin{axis}[
                width=\linewidth,
                height=3cm,
                xlabel={Backdoor Injection Epoch},
                ylabel={SSIM},
                y label style={yshift=-14pt},
                xtick={0, 10, 20, 30, 40, 50},
                xmin=0, xmax=50,
                ymin=0, ymax=1.0,
                tick label style={font=\scriptsize},
                label style={font=\scriptsize},
            ]
            \addplot[blue, thick, line width=1pt] table {
                0  0.1111
                1  0.9197
                3  0.8633
                5  0.8812
                7  0.8984
                9  0.9000
                11 0.8906
                13 0.9546
                15 0.9557
                17 0.9588
                19 0.9348
                21 0.9450
                23 0.9493
                25 0.9429
                27 0.9508
                29 0.9607
                31 0.9612
                33 0.9644
                35 0.9655
                37 0.9657
                39 0.9654
                41 0.9715
                43 0.9803
                45 0.9800
                47 0.9895
                49 0.9897
            };

            \addplot[red, thick, line width=1pt] table {
                0  0.0001
                1  0.0001
                3  0.0001
                5  0.0002
                7  0.0016
                9  0.0035
                11 0.0299
                13 0.0630
                15 0.3932
                17 0.2718
                19 0.4584
                21 0.5838
                23 0.8379
                25 0.9392
                27 0.9479
                29 0.9564
                31 0.9501
                33 0.9523
                35 0.9539
                37 0.9546
                39 0.9554
                41 0.9555
                43 0.9556
                45 0.9563
                47 0.9564
                49 0.9563
            };

            \addplot[green!60!black, thick, line width=1pt] table {
                0  0.0135
                1  0.0001
                3  0.0034
                5  0.0002
                7  0.0128
                9  0.0041
                11 0.0137
                13 0.0092
                15 0.0278
                17 0.0153
                19 0.0377
                21 0.0259
                23 0.4734
                25 0.5466
                27 0.5650
                29 0.4094
                31 0.3996
                33 0.4382
                35 0.6222
                37 0.6971
                39 0.6733
                41 0.6629
                43 0.6952
                45 0.7507
                47 0.7225
                49 0.7230
            };
            \end{axis}
        \end{tikzpicture}
    \end{subfigure}

    \caption{Performance comparison across training epochs for three attacks. Top row: 5\% poison rate. Bottom row: 10\% poison rate.}
    \label{figure:performance_over_time}
\end{figure*}
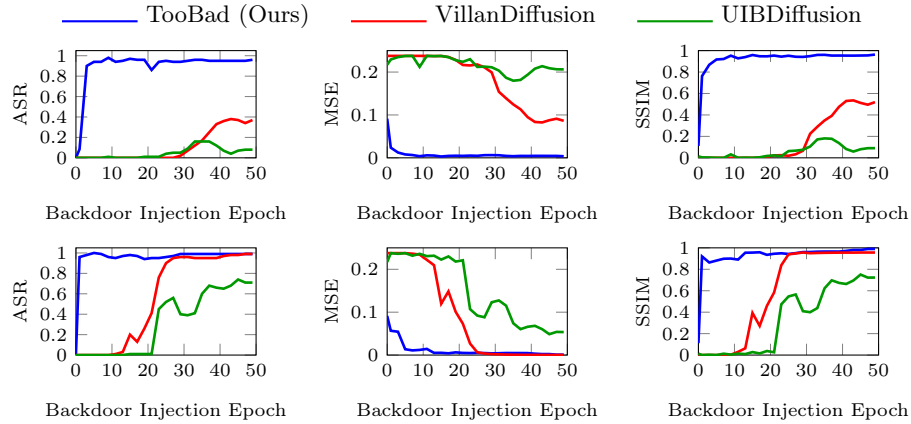

\subsection{Attack Performance Analysis}\label{subsection:experiment-performance} \cref{table:general-performance} presents the performance of TooBad compared to existing baselines across various poison rates. At a 10\% poison rate, TooBad, with almost 100\% ASR, outperforms SOTA methods across all evaluation metrics. The advantage becomes more pronounced at lower poison rates: when the poison rate drops to 5\%, the ASRs of both baselines are reduced dramatically, whereas TooBad still maintains near-perfect attack success with around 98\% ASR, 0.004 MSE, and 0.95 SSIM.
Notably, for poison rates below 5\%, only TooBad remains effective. Both VillanDiffusion and UIBDiffusion fail entirely, resulting in $\sim0$\% ASR. In this case, models backdoored by the baselines only produce black images with arbitrary artifacts, yielding abnormally low MSE and high SSIM. These invalid results are marked as ``X" in \cref{table:general-performance}. 
In contrast, TooBad successfully initiates backdoor behavior at just 0.2\% poison rate and achieves strong performance (93\% ASR, 0.01 MSE, and 0.91 SSIM) at 1\% poison rate.
Some generated examples are visualized in \cref{figure:backdoor-performance-comparison}. 
More generated samples are in the Appendix.

\begin{figure}[t]
\centering

\begin{minipage}[t]{0.52\linewidth}
\centering
\begin{tikzpicture}

\begin{axis}[
    width=0.9\linewidth,
    height=4cm,
    xlabel={Poison Rate (\%)},
    ylabel={MSE},
    xmin=0.2, xmax=1.0,
    ymin=0.01, ymax=0.03,
    xtick={0.2,0.4,0.6,0.8,1.0},
    ytick={0.01,0.015,0.02,0.025,0.03},
    yticklabel style={text=blue},
    y label style={text=blue, at={(axis description cs:0.12,0.3)},anchor=west},
    tick label style={font=\small},
    axis lines=box,
    scaled x ticks=false,
]

\addplot[
    color=blue,
    mark=*,
    thick
] coordinates {
    (0.2,0.030)
    (0.3,0.028)
    (0.4,0.023)
    (0.5,0.021)
    (0.6,0.020)
    (0.7,0.016)
    (0.8,0.015)
    (0.9,0.013)
    (1.0,0.010)
};

\end{axis}

\begin{axis}[
    width=0.9\linewidth,
    height=4cm,
    xmin=0.2, xmax=1.0,
    ymin=0.7, ymax=0.9,
    hide x axis,
    ylabel={SSIM},
    yticklabel style={text=red},
    ytick={0.7,0.75,0.8,0.85,0.9},
    y label style={text=red, at={(axis description cs:1.5,0.3)},anchor=west},
    axis y line*=right,
    axis x line=none,
    tick label style={font=\small},
]

\addplot[
    color=red,
    mark=square*,
    thick
] coordinates {
    (0.2,0.70)
    (0.3,0.766)
    (0.4,0.794)
    (0.5,0.797)
    (0.6,0.807)
    (0.7,0.822)
    (0.8,0.859)
    (0.9,0.877)
    (1.0,0.900)
};

\end{axis}

\end{tikzpicture}

\captionof{figure}{Performance under ultra-low poison rates.}
\label{figure:mse_ssim_super_low_poison}

\end{minipage}
\hfill
\begin{minipage}[t]{0.43\linewidth}
\centering

\begin{tikzpicture}
\begin{axis}[
    width=\linewidth,
    height=3.6cm,
    xlabel={MSE},
    ylabel={FID},
    xmin=0.00, xmax=0.065,
    ymin=4, ymax=12.5,
    xtick={0,0.02,0.04,0.06},
    ytick={5,7,9,11},
    y label style={yshift=-15pt},
    legend style={at={(0.5,1.33)}, anchor=north, legend columns=3, font=\scriptsize},
]

\addplot[
    only marks,
    mark=diamond*, mark size=2pt,
    color=green!60!black
] coordinates {
    (0.043,10.81)
    (0.049,11.02)
    (0.047,9.98)
    (0.046,11.32)
    (0.054,10.12)
};
\addlegendentry{VillanDiff}

\addplot[
    only marks,
    mark=triangle*, mark size=2pt,
    color=red
] coordinates {
    (0.055,10.92)
    (0.042,11.33)
    (0.053,9.55)
    (0.046,9.68)
    (0.050,8.85)
};
\addlegendentry{UIBDiff}

\addplot[
    only marks,
    mark=*,
    mark size=1.8pt,
    color=blue
] coordinates {
    (0.006,7.87)
    (0.014,5.44)
    (0.010,6.42)
    (0.012,6.72)
    (0.013,7.25)
};
\addlegendentry{TooBad}

\end{axis}
\end{tikzpicture}

\captionof{figure}{Utility comparison across different backdoored models.}
\label{figure:fid_comparison}

\end{minipage}

\end{figure}

\subsection{Backdoor under Ultra-Low Poison Rates}\label{subsection:experiment-poison-rate}
This section further highlights the efficiency of our method under extremely low poison rate conditions, ranging from 0.1\% to 1\%. As shown in \cref{figure:ASR-super-low-poison}, only TooBad is able to successfully backdoor DMs in this setting. The ASR of our method rises quickly from 68.4\% at a 0.2\% poison rate to 93\% at 1\%. In contrast, the baseline methods consistently yield 0\% ASR across all tested poison rates.
The corresponding MSE and SSIM scores for TooBad within this range are illustrated in \cref{figure:mse_ssim_super_low_poison}, further validating its effectiveness. Notably, the MSE drops rapidly from 0.03 at a 0.2\% poison rate to only 0.01 at 1\%, while the SSIM score rises steadily from 0.7 to 0.9. These trends indicate that our optimized trigger not only enables successful attacks but also produces backdoored generations that closely resemble the intended target.
In summary, TooBad is the only method that remains effective under extremely low poison rates. This not only makes the attack more practical but also significantly improves its utility.


\subsection{Time and Convergence Analysis}\label{subsection:experiment-time-complexity}
We note that the trigger optimization stage introduces only negligible overhead compared to the backdoor injection stage for two main reasons. First, each trigger optimization iteration runs on a small batch of sampled noises without using any training data, whereas each backdoor injection epoch trains the model on the entire poisoned dataset. Second, backdoor injection fine-tunes the large victim model, while trigger optimization keeps the model frozen and updates only the low-dimensional trigger variables. For example, attacking DDPMs on CIFAR-10, the full process of 50 trigger optimization iterations take about 5 minutes, compared with roughly 2 hours for 50 fine-tuning epochs of backdoor injection.
Consequently, and without compromising fairness, the experiment evaluates performance primarily as a function of the number of backdoor injection epochs.
\cref{figure:performance_over_time} monitors the performance of TooBad and the baselines thorough the backdoor injection process at 5\% and 10\% poison rate. In both cases, our framework converges rapidly, reaching near-perfect performance within just 3-5 backdoor injection epochs. By epoch 5, TooBad already achieves nearly 100\% ASR, close to 1.0 SSIM, and a negligible MSE, while both baselines completely fail to backdoor the victim models. The baselines require more than 30-50 epochs to approach the performance that TooBad attains after only 5 epochs.


\subsection{Stealthiness Analysis}\label{subsection:experiment-stealthiness}

\begin{table*}[htbp]
\centering
\footnotesize
\setlength{\tabcolsep}{5pt}
\caption{Resilience of the attacks against SOTA defenses.}
\begin{tabular}{c|c|c|c|c|c|c}
\hline
Defense & Metric & VillanDiff & UIBDiff & TooBad-NI & TooBad-NS & TooBad \\ \hline
\multirow{4}{*}{Elijah} 
 & ACC & 32.16 & 0.00 & 27.65 & 13.12 & 0.00 \\
 & TPR & 14.77 & 0.00 & 10.23 & 3.34 & 0.00 \\
 & L2D & 38.01 & 40.67 & 38.95 & 39.05 & 41.03 \\ \hline
\multirow{4}{*}{TERD} 
 & ACC & 90.76 & 0.00 & 85.33 & 17.65 & 0.00 \\
 & TPR & 86.23 & 0.00 & 76.66 & 12.03 & 0.00 \\
 & L2D & 27.89 & 40.22 & 32.67 & 38.65 & 40.16 \\ \hline
\multirow{4}{*}{PureDiff} 
 & ACC & 100 & 0.00 & 100 & 19.78 & 0.00 \\
 & TPR & 100 & 0.00 & 100 & 14.07 & 0.00 \\
 & L2D & 24.12 & 41.06 & 25.33 & 39.12 & 41.21 \\ \hline
\end{tabular}
\label{table:backdoor-defense}
\end{table*}

\noindent\textbf{Resistance to SOTA Defenses.} 
We evaluate the robustness of TooBad against three recent defenses: Elijah~\cite{an2024elijah}, TERD~\cite{mo2024terd}, and PureDiffusion~\cite{truong2025dual}. These defenses typically follow a two-stage procedure: first, a trigger inversion stage that try to reconstruct the backdoor trigger from the suspicious model; and second, a detection stage that analyzes the inverted trigger to determine whether the model is backdoored. As shown in \cref{table:backdoor-defense}, TooBad completely evades all three defenses, resulting in a 0\% detection rate across all scenarios. In contrast, the ablated variants TooBad-NS (without sparsity constraint) and TooBad-NI (without invisibility constraint) become detectable. The results show that: (i) the strong resilience of TooBad stems from the imperceptibility constraints applied during trigger optimization, and (ii) the invisibility constraint accounts for the majority of stealthiness. 
Further ablation study for two imperceptibility constraints can be found in the Appendix. 
For the baselines, UIBDiffusion also produces irreversible triggers, while VillanDiffusion is exposed by the defenses.

\noindent\textbf{Utility Evaluation.}
If a backdoored model suffers from low utility, it may fail to generate realistic samples or occasionally produce the backdoor target even without the trigger, making the attack easily detectable. To quantify utility, we primarily use the FID score, where lower values indicate that, in the absence of the trigger, the backdoored model can generate clean samples closely matching the distribution of the original training data. 
In our experiments, each backdoor method is applied to the same set of five pretrained models with varying poison rates (from 0.2\% to 5\%). Since there is an inherent trade-off between utilily and attack performance, both FID and MSE are reported for each resulting backdoored model for fair comparison. As shown in \cref{figure:fid_comparison}, models backdoored by TooBad consistently achieve the lowest FID and lowest MSE, outperforming all baselines in both utility and attack performance. 
These results demonstrate that TooBad not only improve backdoor effectiveness but also preserves the fidelity of clean samples.

\begin{table}[htbp]
\centering
\setlength{\tabcolsep}{5pt}
\footnotesize
\renewcommand{\arraystretch}{0.95}
\caption{Performance on alternative targets beyond the hat image.}
\begin{tabular}{@{}c|ccc|ccc@{}}
\toprule
Target & \multicolumn{3}{c|}{Cat} & \multicolumn{3}{c}{Stop Sign} \\ \midrule
\multicolumn{1}{l|}{Poison Rate} & ASR & MSE & \multicolumn{1}{l|}{SSIM } & ASR & MSE & \multicolumn{1}{l}{SSIM} \\ \midrule
0.2\% & 0.66 & 0.0326 & 0.733 & 0.63 & 0.0368 & 0.633 \\
0.5\% & 0.83 & 0.0112 & 0.883 & 0.81 & 0.0156 & 0.826 \\
1\% & 0.87 & 0.0083 & 0.898 & 0.86 & 0.0096 & 0.869 \\
2\% & 0.92 & 0.0056 & 0.925 & 0.93 & 0.0048 & 0.933 \\
5\% & 0.99 & 0.0022 & 0.981 & 0.99 & 0.0026 & 0.988 \\
10\% & 0.99 & 0.0018 & 0.986 & 0.99 & 0.0021 & 0.978 \\ \bottomrule
\end{tabular}
\label{table:ablation-target}
\end{table}

\subsection{Ablation Study}\label{subsection:ablation}

\noindent\textbf{Alternative Backdoor Targets.}
To demonstrate that the superior performance of TooBad is not dependent on a specific backdoor target, we evaluate it using alternative targets beyond the default fedora hat image. We experiment with varying poison rates from 0.2\% to 10\%, using two new targets: a cat image and a stop-sign image. As shown in \cref{table:ablation-target}, for these targets, TooBad consistently offers strong attack performance across all settings. TooBad begins to successfully backdoor DMs at just 0.2\% poison rate and achieves near-perfect ASR at 5\%, regardless of the chosen target image. These results validate that TooBad can offer superior backdoor performance no matter the chosen backdoor targets, which can be harmful images rather than just cat or hat images.

\begin{table}[htbp]
\centering
\setlength{\tabcolsep}{5pt}
\footnotesize
\renewcommand{\arraystretch}{0.95}
\caption{Performance comparison on NCSNs and CIFAR-10.}
\begin{tabular}{@{}c|cc|cc|cc@{}}
\toprule
Method & \multicolumn{2}{c|}{VillanDiffusion} & \multicolumn{2}{c|}{UIBDiffusion} & \multicolumn{2}{c}{TooBad (Ours)} \\ \midrule
\multicolumn{1}{l|}{Poison rate} & ASR & MSE & ASR & MSE & ASR  & MSE \\ \midrule
25\% & 0 & X & 0 & X & 0.69 & 0.392 \\
30\% & 0 & X & 0 & X & 0.84 & 0.041 \\
35\% & 0 & X & 0 & X & 0.87 & 0.034 \\
40\% & 0 & X & 0 & X & 0.90 & 0.025 \\
45\% & 0 & X & 0 & X & 0.95 & 0.015 \\
50\% & 0.70 & 0.428 & 0.65 & 0.412 & 0.98 & 0.010 \\
70\% & 0.72 & 0.382 & 0.69 & 0.392 & 1.00 & 0.005 \\ \bottomrule
\end{tabular}
\label{table:ablation-ncsn}
\end{table}

\noindent\textbf{Results for NCSNs.}
In addition to DDPMs and LDMs, we evaluate TooBad on NCSNs and compare its performance with baseline methods. 
As noted in~\cite{chou2024villandiffusion}, backdooring NCSNs typically requires significantly higher poison rates than DDPMs to be effective. However, as shown in \cref{table:ablation-ncsn}, TooBad still outperforms existing SOTA methods by a considerable margin. While prior attacks require at least 50\% poison rate to successfully backdoor NCSNs, TooBad achieves comparable at only half that rate, and reaches near-perfect ASR at 50\%. 
Notably, TooBad at 30\% poison rate even achieved higher attack efficiency than both VillanDiffusion and UIBDiffusion at 70\% poison rate. 

\noindent\textbf{Results on CelebA-HQ.} 
\cref{table:ablation-celeb} presents a comparison between TooBad and the baseline methods on the CelebA-HQ dataset using two representative poison rates, 5\% and 10\%. To reduce computational cost, this experiment employs an LDM with a latent space size of $64\times64$. The results demonstrate that TooBad achieves near-perfect ASR at both poison rates, substantially outperforming the baselines. Visualization of generated samples are shown in the Appendix. These findings confirm the effectiveness of our framework on higher-resolution data.

\begin{table}[]
\centering
\setlength{\tabcolsep}{4.2pt}
\caption{Performance comparison on LDMs and CelebA-HQ.}
\footnotesize
\begin{tabular}{@{}l|ccc|ccc@{}}
\toprule
\multicolumn{1}{c|}{Poison rate} & \multicolumn{3}{c|}{$p=5\%$} & \multicolumn{3}{c}{$p=10\%$} \\ \midrule
\multicolumn{1}{c|}{Method} & ASR & MSE & SSIM & ASR & MSE & SSIM \\ \midrule
VillanDiffusion & 0.32 & 0.121 & 0.423 & 0.76 & 0.051 & 0.789 \\
UIBDiffusion & 0.19 & 0.186 & 0.188 & 0.71 & 0.046 & 0.722 \\
TooBad (Ours) & 0.98 & 0.003 & 0.945 & 0.99 & 0.002 & 0.965 \\ \bottomrule
\end{tabular}
\label{table:ablation-celeb}
\end{table}

\section{Conclusion}
We introduced TooBad, a novel backdoor framework that advances the state of backdoor attacks on DMs. Unlike prior methods which struggle with inherent performance trade-offs, TooBad achieves superior attack capability with minimal poison rate and training time. 
It successfully implants backdoors at poison rates less than 1\%, reaching near-perfect ASR at just 5\% poison rate within only a few training epochs.
TooBad also maintains strong stealthiness, high utility, and demonstrates complete resistance to SOTA defense mechanisms. 
Its effectiveness generalizes across different backdoor targets and model types, making it a broadly applicable and practical threat.
These results highlight a critical vulnerability in current generative models and call for urgent development of more robust defenses against such stealthy, low-resource yet highly effective attacks.



%
%
\bibliographystyle{splncs04}
\bibliography{main}

@String(CVPR  = {IEEE Conf. Comput. Vis. Pattern Recog.})

@String(NeurIPS = {Adv. Neural Inform. Process. Syst.})

@String(AAAI  = {AAAI})

@String(CVPR  = {CVPR})

@String(NeurIPS = {NeurIPS})

@inproceedings{ho2020denoising,
  title={Denoising diffusion probabilistic models},
  author={Ho, Jonathan and Jain, Ajay and Abbeel, Pieter},
  booktitle={Advances in Neural Information Processing Systems},
  volume={33},
  pages={6840--6851},
  year={2020}
}

@article{truong2024attacks,
  title={Attacks and defenses for generative diffusion models: A comprehensive survey},
  author={Truong, Vu Tuan and Dang, Luan Ba and Le, Long Bao},
  journal={ACM Computing Surveys},
  volume={57},
  number={8},
  pages={1--44},
  year={2025},
  publisher={ACM New York, NY}
}

@article{heusel2017gans,
  title={Gans trained by a two time-scale update rule converge to a local nash equilibrium},
  author={Heusel, Martin and Ramsauer, Hubert and Unterthiner, Thomas and Nessler, Bernhard and Hochreiter, Sepp},
  journal={Advances in Neural Information Processing Systems},
  volume={30},
  year={2017}
}

@InProceedings{Chou2023CVPR,
    author    = {Chou, Sheng-Yen and Chen, Pin-Yu and Ho, Tsung-Yi},
    title     = {How to Backdoor Diffusion Models?},
    booktitle = {CVPR},
    month     = {June},
    pages     = {4015-4024},
    year      = {2023}
}

@inproceedings{chen2023trojdiff,
  title={Trojdiff: Trojan attacks on diffusion models with diverse targets},
  author={Chen, Weixin and Song, Dawn and Li, Bo},
  booktitle={CVPR},
  pages={4035--4044},
  year={2023}
}

@inproceedings{chou2024villandiffusion,
  title={Villandiffusion: A unified backdoor attack framework for diffusion models},
  author={Chou, Sheng-Yen and Chen, Pin-Yu and Ho, Tsung-Yi},
  booktitle={NeuRIPS},
  pages={33912--33964},
  year={2023}
}

@inproceedings{truong2024text,
  title={Text-guided real-world-to-3D generative models with real-time rendering on mobile devices},
  author={Truong, Vu Tuan and Le, Long Bao},
  booktitle={Proceedings of the IEEE Wireless Communications and Networking Conference},
  pages={1--6},
  year={2024},
  organization={IEEE}
}

@inproceedings{an2024elijah,
  title={Elijah: Eliminating backdoors injected in diffusion models via distribution shift},
  author={An, Shengwei and Chou, Sheng-Yen and Zhang, Kaiyuan and Xu, Qiuling and Tao, Guanhong and Shen, Guangyu and Cheng, Siyuan and Ma, Shiqing and Chen, Pin-Yu and Ho, Tsung-Yi and others},
  booktitle={Proceedings of the AAAI Conference on Artificial Intelligence},
  volume={38},
  pages={10847--10855},
  year={2024}
}

@inproceedings{han2024uibdiffusion,
  title={UIBDiffusion: Universal Imperceptible Backdoor Attack for Diffusion Models},
  author={Han, Yuning and Zhao, Bingyin and Chu, Rui and Luo, Feng and Sikdar, Biplab and Lao, Yingjie},
  booktitle={CVPR},
  pages={19186--19196},
  year={2025}
}

@inproceedings{moosavi2017universal,
  title={Universal adversarial perturbations},
  author={Moosavi-Dezfooli, Seyed-Mohsen and Fawzi, Alhussein and Fawzi, Omar and Frossard, Pascal},
  booktitle={Proceedings of the IEEE Conference on Computer Vision and Pattern Recognition},
  pages={1765--1773},
  year={2017}
}

@inproceedings{zhang2020generalizing,
  title={Generalizing universal adversarial attacks beyond additive perturbations},
  author={Zhang, Yanghao and Ruan, Wenjie and Wang, Fu and Huang, Xiaowei},
  booktitle={Proceedings of the IEEE International Conference on Data Mining},
  pages={1412--1417},
  year={2020},
  organization={IEEE}
}

@article{mo2024terd,
  title={TERD: A unified framework for safeguarding diffusion models against backdoors},
  author={Mo, Yichuan and Huang, Hui and Li, Mingjie and Li, Ang and Wang, Yisen},
  journal={arXiv preprint arXiv:2409.05294},
  year={2024}
}

@inproceedings{rombach2022high,
  title={High-resolution image synthesis with latent diffusion models},
  author={Rombach, Robin and Blattmann, Andreas and Lorenz, Dominik and Esser, Patrick and Ommer, Bj{\"o}rn},
  booktitle={Proceedings of the IEEE/CVF Conference on Computer Vision and Pattern Recognition},
  pages={10684--10695},
  year={2022}
}

@inproceedings{watson2021learning,
  title={Learning fast samplers for diffusion models by differentiating through sample quality},
  author={Watson, Daniel and Chan, William and Ho, Jonathan and Norouzi, Mohammad},
  booktitle={Proceedings of the International Conference on Learning Representations},
  year={2021}
}

@inproceedings{nichol2021glide,
  title={Glide: Towards photorealistic image generation and editing with text-guided diffusion models},
  author={Nichol, Alex and Dhariwal, Prafulla and Ramesh, Aditya and Shyam, Pranav and Mishkin, Pamela and McGrew, Bob and Sutskever, Ilya and Chen, Mark},
  booktitle={Proceedings of the International Conference on Machine Learning},
  pages={16784--16804},
  year={2021}
}

@inproceedings{austin2021structured,
  title={Structured denoising diffusion models in discrete state-spaces},
  author={Austin, Jacob and Johnson, Daniel D and Ho, Jonathan and Tarlow, Daniel and Van Den Berg, Rianne},
  booktitle={Advances in Neural Information Processing Systems},
  volume={34},
  pages={17981--17993},
  year={2021}
}

@inproceedings{hoogeboom2021argmax,
  title={Argmax flows and multinomial diffusion: Learning categorical distributions},
  author={Hoogeboom, Emiel and Nielsen, Didrik and Jaini, Priyank and Forr{\'e}, Patrick and Welling, Max},
  booktitle={Advances in Neural Information Processing Systems},
  volume={34},
  pages={12454--12465},
  year={2021}
}

@inproceedings{li2022diffusion,
  title={Diffusion-lm improves controllable text generation},
  author={Li, Xiang and Thickstun, John and Gulrajani, Ishaan and Liang, Percy S and Hashimoto, Tatsunori B},
  booktitle={Advances in Neural Information Processing Systems},
  volume={35},
  pages={4328--4343},
  year={2022}
}

@inproceedings{chen2020wavegrad,
  title={Wavegrad: Estimating gradients for waveform generation},
  author={Chen, Nanxin and Zhang, Yu and Zen, Heiga and Weiss, Ron J and Norouzi, Mohammad and Chan, William},
  booktitle={Proceedings of the International Conference on Learning Representations},
  year={2020}
}

@inproceedings{popov2021grad,
  title={Grad-tts: A diffusion probabilistic model for text-to-speech},
  author={Popov, Vadim and Vovk, Ivan and Gogoryan, Vladimir and Sadekova, Tasnima and Kudinov, Mikhail},
  booktitle={Proceedings of the International Conference on Machine Learning},
  pages={8599--8608},
  year={2021}
}

@inproceedings{xu2023dream3d,
  title={Dream3d: Zero-shot text-to-3d synthesis using 3d shape prior and text-to-image diffusion models},
  author={Xu, Jiale and Wang, Xintao and Cheng, Weihao and Cao, Yan-Pei and Shan, Ying and Qie, Xiaohu and Gao, Shenghua},
  booktitle={Proceedings of the IEEE/CVF Conference on Computer Vision and Pattern Recognition},
  pages={20908--20918},
  year={2023}
}

@inproceedings{jiang2021talk,
  title={Talk-to-edit: Fine-grained facial editing via dialog},
  author={Jiang, Yuming and Huang, Ziqi and Pan, Xingang and Loy, Chen Change and Liu, Ziwei},
  booktitle={Proceedings of the IEEE/CVF International Conference on Computer Vision},
  pages={13799--13808},
  year={2021}
}

@inproceedings{xu2022geodiff,
  title={Geodiff: A geometric diffusion model for molecular conformation generation},
  author={Xu, Minkai and Yu, Lantao and Song, Yang and Shi, Chence and Ermon, Stefano and Tang, Jian},
  booktitle={Proceedings of the International Conference on Learning Representations},
  year={2021}
}

@inproceedings{luo2022antigen,
  title={Antigen-specific antibody design and optimization with diffusion-based generative models for protein structures},
  author={Luo, Shitong and Su, Yufeng and Peng, Xingang and Wang, Sheng and Peng, Jian and Ma, Jianzhu},
  booktitle={Advances in Neural Information Processing Systems},
  pages={9754--9767},
  year={2022}
}

@inproceedings{tashiro2021csdi,
  title={Csdi: Conditional score-based diffusion models for probabilistic time series imputation},
  author={Tashiro, Yusuke and Song, Jiaming and Song, Yang and Ermon, Stefano},
  booktitle={Advances in Neural Information Processing Systems},
  volume={34},
  pages={24804--24816},
  year={2021}
}

@article{yan2021scoregrad,
  title={Scoregrad: Multivariate probabilistic time series forecasting with continuous energy-based generative models},
  author={Yan, Tijin and Zhang, Hongwei and Zhou, Tong and Zhan, Yufeng and Xia, Yuanqing},
  journal={arXiv preprint arXiv:2106.10121},
  year={2021}
}

@inproceedings{rasul2020multivariate,
  title={Multivariate probabilistic time series forecasting via conditioned normalizing flows},
  author={Rasul, Kashif and Sheikh, Abdul-Saboor and Schuster, Ingmar and Bergmann, Urs and Vollgraf, Roland},
  booktitle={Proceedings of the International Conference on Learning Representations},
  year={2020}
}

@inproceedings{kingma2014vae,
  title={Auto-Encoding Variational Bayes},
  author={Kingma, Diederik P and Welling, Max},
  booktitle={Proceedings of the International Conference on Machine Learning},
  year={2014}
}

@inproceedings{rezende2015variational,
  title={Variational inference with normalizing flows},
  author={Rezende, Danilo and Mohamed, Shakir},
  booktitle={Proceedings of the International Conference on Machine Learning},
  pages={1530--1538},
  year={2015}
}

@inproceedings{goodfellow2014generative,
  title={Generative adversarial nets},
  author={Goodfellow, Ian and Pouget-Abadie, Jean and Mirza, Mehdi and Xu, Bing and Warde-Farley, David and Ozair, Sherjil and Courville, Aaron and Bengio, Yoshua},
  booktitle={Advances in Neural Information Processing Systems},
  volume={27},
  year={2014}
}

@inproceedings{song2020score,
  title={Score-based generative modeling through stochastic differential equations},
  author={Song, Yang and Sohl-Dickstein, Jascha and Kingma, Diederik P and Kumar, Abhishek and Ermon, Stefano and Poole, Ben},
  booktitle={Proceedings of the International Conference on Learning Representations},
  year={2021}
}

@inproceedings{song2020denoising,
  title={Denoising diffusion implicit models},
  author={Song, Jiaming and Meng, Chenlin and Ermon, Stefano},
  booktitle={Proceedings of the International Conference on Learning Representations},
  year={2021}
}

@inproceedings{song2019generative,
  title={Generative modeling by estimating gradients of the data distribution},
  author={Song, Yang and Ermon, Stefano},
  booktitle={Advances in Neural Information Processing Systems},
  volume={32},
  pages={11918--–11930},
  year={2019}
}

@inproceedings{struppek2023rickrolling,
  title={Rickrolling the artist: Injecting backdoors into text encoders for text-to-image synthesis},
  author={Struppek, Lukas and Hintersdorf, Dominik and Kersting, Kristian},
  booktitle={Proceedings of the IEEE/CVF International Conference on Computer Vision},
  pages={4584--4596},
  year={2023}
}

@inproceedings{wang2023stronger,
  title={The stronger the diffusion model, the easier the backdoor: Data poisoning to induce copyright breaches without adjusting finetuning pipeline},
  author={Wang, Haonan and Shen, Qianli and Tong, Yao and Zhang, Yang and Kawaguchi, Kenji},
  booktitle={Advances in Neural Information Processing Systems},
  booksubtitle = {Workshop on Backdoors in Deep Learning},
  year={2023}
}

@inproceedings{pan2023trojan,
  title={From Trojan Horses to Castle Walls: Unveiling Bilateral Backdoor Effects in Diffusion Models},
  author={Pan, Zhuoshi and Yao, Yuguang and Liu, Gaowen and Shen, Bingquan and Zhao, H Vicky and Kompella, Ramana Rao and Liu, Sijia},
  booktitle={Advances in Neural Information Processing Systems},
  booksubtitle = {Workshop on BUGS},
  year={2023}
}

@inproceedings{zhai2023text,
  title={Text-to-image diffusion models can be easily backdoored through multimodal data poisoning},
  author={Zhai, Shengfang and Dong, Yinpeng and Shen, Qingni and Pu, Shi and Fang, Yuejian and Su, Hang},
  booktitle={ACM Multimedia},
  pages={1577--1587},
  year={2023}
}

@article{truong2024purediffusion,
  title={PureDiffusion: Using Backdoor to Counter Backdoor in Generative Diffusion Models},
  author={Truong, Vu Tuan and Le, Long Bao},
  journal={arXiv preprint arXiv:2409.13945},
  year={2024}
}

@article{truong2025dual,
  title={A Dual-Purpose Framework for Backdoor Defense and Backdoor Amplification in Diffusion Models},
  author={Truong, Vu Tuan and Le, Long Bao},
  journal={arXiv preprint arXiv:2502.19047},
  year={2025}
}

@article{madry2017towards,
  title={Towards deep learning models resistant to adversarial attacks},
  author={Madry, Aleksander and Makelov, Aleksandar and Schmidt, Ludwig and Tsipras, Dimitris and Vladu, Adrian},
  journal={arXiv preprint arXiv:1706.06083},
  year={2017}
}

@article{krizhevsky2009learning,
  title={Learning multiple layers of features from tiny images},
  author={Krizhevsky, Alex and Hinton, Geoffrey and others},
  year={2009}
}

@inproceedings{liu2015deep,
  title={Deep learning face attributes in the wild},
  author={Liu, Ziwei and Luo, Ping and Wang, Xiaogang and Tang, Xiaoou},
  booktitle={Proceedings of the IEEE/CVF International Conference on Computer Vision},
  pages={3730--3738},
  year={2015}
}

@article{yang2023diffusion,
  title={Diffusion models: A comprehensive survey of methods and applications},
  author={Yang, Ling and Zhang, Zhilong and Song, Yang and Hong, Shenda and Xu, Runsheng and Zhao, Yue and Zhang, Wentao and Cui, Bin and Yang, Ming-Hsuan},
  journal={ACM Computing Surveys},
  volume={56},
  number={4},
  pages={1--39},
  year={2023}
}

@ARTICLE{10419041,
  author={Cao, Hanqun and Tan, Cheng and Gao, Zhangyang and Xu, Yilun and Chen, Guangyong and Heng, Pheng-Ann and Li, Stan Z.},
  journal={IEEE Transactions on Knowledge and Data Engineering}, 
  title={A Survey on Generative Diffusion Models}, 
  year={2024},
  volume={},
  number={},
  pages={1-20},
  doi={10.1109/TKDE.2024.3361474}}

@ARTICLE{10081412,
  author={Croitoru, Florinel-Alin and Hondru, Vlad and Ionescu, Radu Tudor and Shah, Mubarak},
  journal={IEEE Transactions on Pattern Analysis and Machine Intelligence}, 
  title={Diffusion Models in Vision: A Survey}, 
  year={2023},
  volume={45},
  number={9},
  pages={10850-10869},
  doi={10.1109/TPAMI.2023.3261988}}

@article{zou2023diffusion,
  title={Diffusion models in nlp: A survey},
  author={Zou, Hao and Kim, Zae Myung and Kang, Dongyeop},
  journal={arXiv preprint arXiv:2305.14671},
  year={2023}
}

@article{li2024invisible,
  title={Invisible backdoor attacks on diffusion models},
  author={Li, Sen and Ma, Junchi and Cheng, Minhao},
  journal={arXiv preprint arXiv:2406.00816},
  year={2024}
}

@ARTICLE{10552303,
  author={Gao, Yinghua and Li, Yiming and Gong, Xueluan and Li, Zhifeng and Xia, Shu-Tao and Wang, Qian},
  journal={IEEE Transactions on Information Forensics and Security}, 
  title={Backdoor Attack With Sparse and Invisible Trigger}, 
  year={2024},
  volume={19},
  number={},
  pages={6364-6376},
  doi={10.1109/TIFS.2024.3411936},
  ISSN={1556-6021},
  month={},}

@inproceedings{saha2020hidden,
  title={Hidden trigger backdoor attacks},
  author={Saha, Aniruddha and Subramanya, Akshayvarun and Pirsiavash, Hamed},
  booktitle={Proceedings of the AAAI conference on artificial intelligence},
  volume={34},
  number={07},
  pages={11957--11965},
  year={2020}
}

@inproceedings{liang2024badclip,
  title={Badclip: Dual-embedding guided backdoor attack on multimodal contrastive learning},
  author={Liang, Siyuan and Zhu, Mingli and Liu, Aishan and Wu, Baoyuan and Cao, Xiaochun and Chang, Ee-Chien},
  booktitle={Proceedings of the IEEE/CVF Conference on Computer Vision and Pattern Recognition},
  pages={24645--24654},
  year={2024}
}

@article{vltrojan,
  title={Vl-trojan: Multimodal instruction backdoor attacks against autoregressive visual language models},
  author={Liang, Jiawei and Liang, Siyuan and Liu, Aishan and Cao, Xiaochun},
  journal={Proceedings of the International Journal of Computer Vision},
  pages={1--20},
  year={2025},
  publisher={Springer}
}

@inproceedings{ngiam2011learning,
  title={Learning deep energy models},
  author={Ngiam, Jiquan and Chen, Zhenghao and Koh, Pang W and Ng, Andrew Y},
  booktitle={Proceedings of the International Conference on Machine Learning},
  pages={1105--1112},
  year={2011}
}

@article{wang2004image,
  title={Image quality assessment: from error visibility to structural similarity},
  author={Wang, Zhou and Bovik, Alan C and Sheikh, Hamid R and Simoncelli, Eero P},
  journal={IEEE Transactions on Image Processing},
  volume={13},
  number={4},
  pages={600--612},
  year={2004},
  publisher={IEEE}
}

@article{gu2017badnets,
  title={Badnets: Identifying vulnerabilities in the machine learning model supply chain},
  author={Gu, Tianyu and Dolan-Gavitt, Brendan and Garg, Siddharth},
  journal={arXiv preprint arXiv:1708.06733},
  year={2017}
}

@article{blended,
  title={Targeted backdoor attacks on deep learning systems using data poisoning},
  author={Chen, Xinyun and Liu, Chang and Li, Bo and Lu, Kimberly and Song, Dawn},
  journal={arXiv preprint arXiv:1712.05526},
  year={2017}
}

@inproceedings{trojanvqa,
  title={Dual-key multimodal backdoors for visual question answering},
  author={Walmer, Matthew and Sikka, Karan and Sur, Indranil and Shrivastava, Abhinav and Jha, Susmit},
  booktitle={Proceedings of the IEEE/CVF Conference on Computer Vision and Pattern Recognition},
  pages={15375--15385},
  year={2022}
}

@inproceedings{yang2022transferable,
  title={Transferable graph backdoor attack},
  author={Yang, Shuiqiao and Doan, Bao Gia and Montague, Paul and De Vel, Olivier and Abraham, Tamas and Camtepe, Seyit and Ranasinghe, Damith C and Kanhere, Salil S},
  booktitle={Proceedings of the International Symposium on Research in Sttacks, Intrusions and Defenses},
  pages={321--332},
  year={2022}
}

@inproceedings{he2016deep,
  title={Deep residual learning for image recognition},
  author={He, Kaiming and Zhang, Xiangyu and Ren, Shaoqing and Sun, Jian},
  booktitle={Proceedings of the IEEE/CVF Conference on Computer Vision and Pattern Recognition},
  pages={770--778},
  year={2016}
}

@inproceedings{jiang2023color,
  title={Color backdoor: A robust poisoning attack in color space},
  author={Jiang, Wenbo and Li, Hongwei and Xu, Guowen and Zhang, Tianwei},
  booktitle={Proceedings of the IEEE/CVF conference on computer vision and pattern recognition},
  pages={8133--8142},
  year={2023}
}
\end{document}